# Effect of cooperative merging on the synchronous flow phase of traffic

L. C. Davis*, Physics Dept., University of Michigan, Ann Arbor, MI 48109

**ABSTRACT**

Cooperation in merging is introduced by adding interactions between pairs of vehicles in opposite lanes. Simulations with an improved version of the modified optimal velocity model are done for two lanes merging into a single lane. For ~30 seconds prior to reaching the merge region, vehicles in both lanes adjust their headways to create safe distances in front of and behind the merging vehicle. Cooperation prevents the transition from free flow to synchronous flow that occurs for normal merging, provided the merge region is sufficiently large and the total incoming flow does not exceed the maximum possible single-lane flow. No long-range vehicle-to-vehicle communication is required for the type of cooperation considered.

* Email: ldavis7@peoplepc.com

I. **INTRODUCTION**

Synchronous flow (SF), like a traffic jam, is an emergent phenomenon [1]. Drivers unintentionally self organize to transform free flow (FF) into a new phase [2]. In most instances, the transition is initially FF to SF and then later it can be followed by SF to a jam, although FF to a jam is possible, according to the three-phase model of Kerner and coworkers [3]. These transitions are considered first order. For reviews of traffic modeling and empirical data see reference [4].

Synchronous flow, a phase of traffic that can extend over macroscopic distances (> 1 km), generally results from vehicles' merging (from an on-ramp) with the incoming flow on the main road. Local rules for merging, such as requiring speed-dependent safe distances to the preceding and following vehicles in the opposite lane, are sufficient to produce SF in simulations using realistic vehicle dynamics when the flow is high enough [5].

The probability to form jams can be reduced by anticipation, that is, making use of information about more than just the closest vehicle in front [6]. Looking ahead (observing tail lights for several preceding vehicles in the same lane, for example) helps drivers anticipate the response they should make. It has been found that anticipation stabilizes flow against the formation of jams [6]. It is interesting to ask if there exist analogous changes in the local merging rules that suppress the FF to SF transition. The purpose of this paper is to present one such change involving cooperation of pairs of drivers (vehicles) [7] and to evaluate its effects. This cooperation between pairs is required over times of approximately 30 seconds (hereafter denoted by s), but does not require long-range vehicle-vehicle communication. The study of cooperative merging is important because it improves our understanding of SF formation; and it could have implications for improving traffic flow and reducing congestion.

A schematic diagram of the vehicle interactions is shown in Fig. 1. Within a distance $|z_0|$ of the end of the merge region, a vehicle on the on-ramp (denoted by B) not only interacts with the vehicle immediately preceding it in the same lane (denoted by 2), but also with a vehicle, if it is closer, in the opposite lane (denoted by A). Likewise, a vehicle on the main road (denoted by 1) interacts with the preceding vehicle in the same lane (A) and the vehicle on the on-ramp (B), if it is closer. The purpose of the additional interactions is to ensure that a safe distance exists in front of and behind the merging vehicle when it reaches the merge region ($-d_{merge} < x < 0$).

The paper is organized as follows. The model is presented in Section II. Results from simulations using the model are given in Section III. The summary is given in Section IV.

## II. TRAFFIC MODEL FOR SIMULATIONS

The dynamical model for vehicular motion is described in this section. The model improves upon the modified optimal velocity model [5] primarily by including mechanical constraints on acceleration and deceleration [8]. The model also incorporates a velocity-dependent synchronization distance and is fully consistent with the three-phase model [3]. In addition, the rules for merging [5] have been altered slightly to be more self-consistent. Randomly distributed vehicle time constants and a power-law headway distribution for the initial conditions are used. These improvements make the traffic model more realistic.

**The Model**

The position and velocity of the $n^{th}$ vehicle as a function of time $t$ are denoted by $x_n(t)$ and $v_n(t)$. Ascending values of the index $n$ correspond to increasingly negative initial positions $x_n(0)$ in each lane (main line and on-ramp). The delay time due to driver reaction is $t_d$. Let the effective headway (including vehicle length) be

$$\Delta_n(t) = \Delta x_n(t - t_d) + t_d \Delta v_n(t - t_d), \tag{1}$$

where

$$\Delta x_n(t) = x_{n-1}(t) - x_n(t) \tag{2}$$

and

$$\Delta v_n(t) = v_{n-1}(t) - v_n(t). \tag{3}$$

Vehicle *n-1* is immediately in front of vehicle *n* in the same lane. The equation of motion for the *n*th vehicle is

$$\tau_n \frac{dv_n(t)}{dt} + v_n(t) = V_{desired}(t), \tag{4}$$

where

$$V_{desired}(t) = V_{OV}(\Delta_n(t)) \tag{5}$$

if $V_{OV}(\Delta_n(t)) < v_n(t)$, where $V_{OV}$ is the optimal velocity function [9]. To stabilize vehicle motion (to eliminate wild oscillations in velocity, for example [10]), it is necessary to replace $V_{OV}$ by the velocity of the preceding vehicle under some conditions [5]. For $V_{OV}(\Delta_n(t)) \geq v_n(t)$

$$V_{desired}(t) = \min\{V_{OV}(\Delta_n(t)), \quad v_{n-1}(t - t_d)\}, \tag{6}$$

where the optimal velocity function is [9]

$$V_{OV}(h) = V_0\{\tanh[C_1(h - h^0)] + C_2\}. \tag{7}$$

Eq. (6) holds if the headway is small enough, that is, if $\Delta_n(t) < 2H_{OV}(v_{n-1}(t - t_d))$, where the inverse function $H_{OV}$ is defined by

$$H_{OV}(V_{OV}(h)) = h. \tag{8}$$

$H_{OV}$ is the equilibrium headway (in the optimal velocity model of Bando *et al.* [9]) at a given velocity. For larger headways, the model is modified to allow lagging vehicles to catch up. When $\Delta_n(t) \geq 2H_{OV}(v_{n-1}(t - t_d))$ and $V_{OV}(\Delta_n(t)) \geq v_n(t)$, we have

$$V_{desired}(t) = V_{OV}(\Delta_n(t)) + [v_{n-1}(t - t_d) - V_{OV}(\Delta_n(t))]\exp\left(1 - \frac{\Delta_n(t)}{2H_{OV}(v_{n-1}(t - t_d))}\right). \tag{9}$$

Note that $2H_{OV}(v_{n-1}(t-t_d))$ in Eq. (9) replaces the fixed distance $L$ of reference 5. $L$ was introduced originally [5] to cause vehicles to close a large gap to the preceding vehicle and it is more realistic to make it velocity dependent than fixed. The constants in Eq. (7) are given by $C_1 = 0.086$/m, $C_2 = 0.913$, $h^0 = 25$ m, and $V_0 = 16.8$ m/s [11].

Note that there are solutions to this model of the form $v_n(t) = v_0$ and $\Delta x_n(t) = \Delta x_n(0)$ where the initial differences in position satisfy $H_{OV}(v_0) \leq \Delta x_n(0) \leq 2 H_{OV}(v_0)$ but are otherwise arbitrary. Thus the model satisfies the basic postulate of the three-phase model [3], namely that equilibrium solutions can occupy a two-dimensional region of flow-density space.

The constraints imposed by vehicle mechanical limitations are

$$a_{accel} dt \geq dv_n(t) \geq -a_{decel} dt . \tag{10}$$

Here the maximum acceleration is $a_{accel} = 3$ m/s$^2$ and the maximum deceleration is

$a_{decel} = 10$ m/s$^2$.

To avoid collisions, the Gipps-like condition [12]

$$dv_n(t) \leq -a_g dt \tag{11}$$

must be satisfied when

$$\Delta x_n(t-t_d) + \frac{v_{n-1}^2(t-t_d) - v_n^2(t-t_d)}{2a_g} - t_d v_n(t-t_d) < D . \tag{12}$$

Here $D = 7$ m and $a_g$ is 3 m/s$^2$. The constraint $dv_n(t) = -a_g dt$ is imposed when the equation of motion fails to give sufficient deceleration.

For either lane, a speed limit is imposed so that

$$V_{desired}(t) \le v_{Limit}. \tag{13}$$

Operationally, if $V_{desired}$ becomes larger than $v_{Limit}$, it is replaced by the speed limit.

**Rules for Merging**

The region for vehicles in lane 2 to merge into lane 1 is of length $d_{merge}$. If at time $t$ the vehicle in lane 2 selected to merge is $n$ and

$$-d_{merge} < x_n(t - t_d) < 0, \tag{14}$$

$n$ is permitted to change lanes when the following conditions are satisfied. Let $nf$ ($nb$) be the vehicle in lane 1 directly in front of (behind) $n$. Then

$$x_{nf}(t - t_d) - x_n(t - t_d) > S_f H_{OV}(v_n(t - t_d)) \tag{15}$$

and

$$x_n(t - t_d) - x_{nb}(t - t_d) > S_f H_{OV}(v_{nb}(t - t_d)) \tag{16}$$

must be satisfied. Here $H_{OV}$ serves as a velocity-dependent safe distance for merging. The factor $S_f$ is taken to be 0.7. By trial and error, this value was found to give satisfactory merging that did not interrupt mainline flow substantially, yet permitted reasonable merge rates and merge velocities. Vehicles are selected at random to possibly merge every 0.05 s.

The lead vehicle in lane 2 approaches the downstream end of the merge region as if a phantom vehicle existed at $x = 0$ with $v = v_{Limit}$. In addition, if

$$x_n(t - t_d) > -\frac{v_n^2(t - t_d)}{a_g}, \tag{17}$$

then

$$dv_n(t) = -a_g dt. \tag{18}$$

Note that a factor of 2 is omitted in the denominator on the RHS to prevent significant travel beyond $x = 0$, the downstream end of the merge region.

**Power-law Distribution of Initial Headways**

To mimic approximately, but simply, empirical headways [13] a power-law distribution is employed for specifying the initial conditions. The headways are defined here as the distance from the center of one vehicle to the center of the next. A sequence of headways is generated by repeatedly setting

$$\left(\frac{h_0}{h}\right)^n = r, \tag{19}$$

where $r$ is a random number $0 \leq r \leq 1$ and $h_0$ is the smallest headway. It is determined by the initial velocity $v_0$ according to $h_0 = H_{OV}(v_0)$. The average headway is

$$\bar{h} = \frac{n}{n-1} h_0. \tag{20}$$

The probability that a headway is in the range $h$ to $h + \delta h$ is $P(h)\, \delta h$ for $h \geq h_0$ where

$$P(h) = \frac{n}{h_0}\left(\frac{h_0}{h}\right)^{n+1}. \tag{21}$$

The probability vanishes for $h < h_0$.

**Cooperative Merging Formalism**

In this section, I formulate the way a vehicle gradually adjusts its position to the preceding vehicle in opposite lane as they approach the merge region. The objective of cooperative merging is to have safe headways in front and back so that the merging vehicle can change lanes without slowing down appreciably. Let $z_0$ be a point upstream of the merge region. Consider a vehicle (denoted by 1 in Fig. 1) in lane 1 with label *n*. Then for $z_0 < x_n(t-t_d) < -d_{merge}$ let

$$\alpha = 1 - \frac{x_n(t-t_d) + d_{merge}}{z_0 + d_{merge}} \qquad (22)$$

and for $-d_{merge} < x_n(t-t_d) < 0$

$$\alpha = 1. \qquad (23)$$

Let $x^B(t-t_d)$ be the position of the nearest preceding vehicle (denoted by B in Fig. 1), which I take to be in lane 2. Let [See. Eq. (7).]

$$V^B(t) = 0.99 V_{OV}(x^B(t-t_d) - x_n(t-t_d)) \qquad (24)$$

[The factor of 0.99 is included to provide a small margin of error.] If $V^B(t) < V_{desired}(t)$, then $V_{desired}(t)$ [See Eq. (6).] is replaced by

$$\tilde{V}_{desired}(t) = \alpha V^B(t) + (1-\alpha) V_{desired}(t). \qquad (25)$$

This part of cooperative merging ensures there is a suitable gap on the main line behind the merging vehicle. If $x^B(t-t_d) > x_{n-1}(t-t_d)$ [*n-1* is the preceding vehicle in lane 1, denoted by A in Fig.1] then I set $\alpha = 0$, because the main line vehicle should not come too close to the preceding vehicle in the same lane.

The other part of cooperation involves the on-ramp vehicles. For a vehicle in lane 2 (denoted by B in Fig. 1) with label *n*, let $x^A(t-t_d)$ be the position of the nearest preceding vehicle in lane 1 (denoted by A in Fig. 1). Define $V^A(t)$ analogously to Eq. (24) with $x^B$ replaced by $x^A$ and $\tilde{V}_{desired}(t)$ by Eq. (25) with $V^B$ replaced by $V^A$. For any vehicle in lane 2, *other than the first*, if $x^A(t-t_d) > x_{n-1}(t-t_d)$ [*n-1* is now in lane 2 and denoted by 2 in Fig. 1] then $\alpha = 0$. This interaction produces a safe gap on the main line in front of the merging vehicle.

For either lane, I require

$$\tilde{V}_{desired}(t) \leq v_{Limit}. \tag{26}$$

If $\tilde{V}_{desired}(t) > v_{Limit}$ then it is replaced by the speed limit.

## III. RESULTS

In this section I present results from simulations using the model described in Sec. II. For a power-law distribution with $n = 3$ and $h_0 = 50$ m, a typical distribution is shown in Fig. 2. Note that $n = 3$ corresponds to $P(h) \propto 1/h^4$. The distribution is similar to that observed [11]. For the calculations presented here the *possible* initial positions of vehicles are determined by letting $x_k = x_{k-1} - h_k$ where the $k^{th}$ headway $h_k$ is given by the $k^{th}$ random number in Eq. (19). [For lane 1, $x_0 = 0$, the initial position of the lead vehicle.] However, not all possible initial positions are necessarily occupied. I randomly occupy sites in lane 1 (2) with probability $p_1$ ($p_2$). Fig. 2 pertains to the probability being unity.

Vehicles in lane 2 are initially offset upstream by 1000 m. I take the mechanical time constant $\tau_n$ (in seconds) to be randomly chosen from the interval [0.5,1.0]. The delay time is $t_d = 0.75$ s [5, 10]. A speed limit of 32 m/s is imposed and the length of the merge region is 300 m, unless noted otherwise.

The velocity at which vehicles merge is not only determined by $p_1$ and $p_2$ but also by the length $d_{merge}$. The latter effect is due to deceleration as the lead vehicle in lane 2 approaches the end of the merge region at $x = 0$. From Eq. (17) we can see that maximum merge velocity is approximately $\sqrt{a_g d_{merge}}$. Dynamical effects prevent this from being an exact maximum.

**Simulations**

I take $p_1 = 0.7$, $p_2 = 0.45$, $n = 3$, $h_0 = 50$ m, $d_{merge} = 300$ m, and $z_0 = -1000$ m. In Fig. 3, the position and velocity of merges for cooperative and normal merging are compared. We see that for cooperative merging the merge velocities are between 25 and 30 m/s, whereas normal merging produces a range of values down to 9 m/s. These merges are distributed along the curve $x = -v^2/a_g$ where $a_g = 3$ m/s$^2$ because of deceleration near the downstream end of lane 2 [Eqs. (17) and (18)]. With cooperative merging, vehicles in lane 2 can change lanes almost immediately (recall there is a 0.75-s delay for driver reaction) after entering the merge region because the interacting vehicles adjust headways in the region $z_0 > x > 0$. The effect on the formation of SF of higher merge velocities due to cooperative merging is shown in Fig. 4. With normal merging, a well-defined region of SF forms in lane 1 and extends almost to $x = -1.5$ km by $t = 500$ s. In contrast, vehicles in lane 1 operating with cooperative merging maintain FF throughout this region with only a slight decrease in velocity.

Increasing $p_2$ up to 0.7 (Fig. 5) does not induce SF if merging is cooperative. The total number of vehicles exiting and the number of merges in 500 s increase linearly and the difference between the two remains nearly constant as a function $p_2$. Because merging takes place at reasonably large velocities, vehicles entering from lane 2 do not appreciably decrease the incoming flow from lane 1. Fig. 6 shows the striking difference between the behaviors of the difference (total less merges) as a function of $p_2$. Normal merging produces a marked decrease above $p_2 = 0.4$ where there is a transition to the SF phase in lane 1, while little change occurs for cooperative merging. When SF forms with normal merging, the incoming flow is diminished considerably relative to the upstream flow.

There are, however, limitations to the extent which cooperation suppresses SF formation. If $d_{merge}$ is only 100 m, deceleration due to the approaching end of the merge region limits the merge velocity to about 15 m/s for cooperative merging. In Fig. 7a, the velocity and position of merges are shown for cooperative and normal merging. Here I consider somewhat different parameters, $p_1 = 1.0$, $p_2 = 0.2$, $n = 3$, $h_0 = 40$ m, and a speed limit of 30 m/s. Cooperative merging does not suppress the transition to SF in lane 1 in this case, although it does lessen the effect. (See Fig. 7b.) The average velocity in the congested region is ~ 5 m/s higher with cooperative merging relative to normal merging. However, if $d_{merge}$ is increased to 300 m, the transition to SF is avoided (Fig. 8). For normal merging, SF still occurs as it did for shorter $d_{merge}$. Since the short merge region reduces the maximum velocity at which merges can occur, even if cooperation exists, some reduction of the flow in lane 1 results.

Another example where cooperation does not prevent a transition to a congested phase, but mitigates the effects is shown in Fig. 9. The incoming flow has $p_1 = p_2 = 1$ for $h_0 = 50$ m, which gives a rate of $32/75 = 0.427$/s in each lane. The sum, $0.853$/s, exceeds the theoretical capacity of the outgoing lane, which is $0.772$/s (determined by the maximum value of $V_{OV}(h)/h$ [See Eq. (7).]). The flow rate beyond the merge region for normal merging is about $0.36$/s, which is even less than the incoming flow in either lane. For cooperative merging, the rate fluctuates around $0.7$/s, except for a dip near 400 s caused by a brief period when merging temporarily occurs at small velocities. In this case and the previous one, the formation of SF cannot be avoided because of a constraint (maximum flow limitation or maximum merge velocity due to merge length). If there are no such constraints, cooperation prevents the self-organized congestion.

Physically, cooperative merging suppresses the transition to the synchronous flow phase by ensuring that a merging vehicle has sufficient headway in front and back so that it can change lanes without slowing down or forcing other vehicles to decelerate. In normal merging, a mainline vehicle does not adjust its velocity until a merge occurs in front of it. Up until then it has followed the vehicle immediately ahead in the same lane. When a vehicle merges normally, it often does so at low velocity because it has neared the downstream end of the merge region. This causes some incoming mainline vehicles to decelerate abruptly and congestion can ensue in heavy traffic if there are frequent low-velocity merges. In other situations, the merging vehicle may have to decelerate if the mainline vehicle in front is moving slowly. In either case, cooperation results in a gradual opening of a safe gap for the merging vehicle to fit into prior to reaching the merge region. It does so without inducing congestion. Limits on acceleration and deceleration, not imposed in reference [5], exacerbate the effects of merging and make the effects of cooperation more pronounced.

## IV. SUMMARY

Refinements to the modified optimal velocity model, which take account of the mechanical restrictions on vehicle acceleration and deceleration, have been implemented. Changes to the rules for merging at an on-ramp have been made to introduce cooperation. Here a vehicle on the on-ramp not only interacts with the vehicle in front of it in the same lane (or the end of the merge region if it is the lead vehicle) but also with a vehicle in the opposite lane (main road) when closer. Likewise, a vehicle on the main road interacts with the vehicle in front of it and with a potential merging vehicle if it is closer. The purpose of the additional interactions, which are of a car-following nature, is to adjust headways so that safe distances in front of and behind the merging vehicle are obtained before reaching the region where merging is allowed.

Compared to normal merging, where there is no cooperation, the effects of cooperative merging are striking. Simulations described in Sec. III demonstrate that the transition to the SF phase can be completely suppressed if the merge region is large enough. Of course, the total incoming flow must not exceed the maximum theoretical capacity of the single lane downstream of the on-ramp. If the merge region is short, then vehicles must necessarily merge at low velocities because of the approaching end of the merge region. In this instance, cooperative merging cannot eliminate the transition to SF but it does significantly decrease the time required to traverse the congested region.

The results of this work provide another example where changes in local rules affect the emergent behavior of traffic. Anticipation stabilizes traffic against formation of jams [6] and cooperation suppresses formation of synchronous flow, the two principal phases of congested traffic [2,3]. Anticipation and the additional interaction for lane-2 vehicles in cooperative merging (Fig. 1) can be thought of as actions that directly benefit the driver performing them, whereas the additional interaction for lane-1 vehicles is more altruistic.

**FIGURE CAPTIONS**

Fig. 1 (Color online) Schematic of vehicle interactions for cooperative merging. In the region $z_0 < x < 0$, drivers in both lanes adjust their speed and headway according to an algorithm that accounts for the preceding vehicle in the same lane as well as any closer vehicle in the opposite lane. Vehicle 1 interacts with B in lane 2 and A in lane 1. Vehicle B on the on-ramp interacts with A and 2.

Fig. 2. A typical distribution of headways for $n = 3$ and $h_0 = 50$ m.

Fig. 3. (Color online) Position and velocity of merges for cooperative (squares) and normal merging (diamonds). The solid line is $x = -v^2 / a_g$ where $a_g = 3$ m/s$^2$. The parameters are $p_1 = 0.7$, $p_2 = 0.45$, $n = 3$, $h_0 = 50$ m with a lane-2 offset of 1000 m, $d_{merge} = 300$ m, and $z_0 = -1000$ m. The limiting velocity is 32 m/s. The delay time is $t_d = 0.75$ s and $0.5$ s $< \tau_n < 1.0$ s.

Fig. 4. (Color online) Velocity *vs.* position of vehicles in lane 1 at $t = 500$ s for cooperative (squares) and normal merging (diamonds). The parameters are same as in Fig. 3.

Fig. 5. (Color online) The total number of vehicles exiting in 500 s (diamonds), the number of merges (squares), and the difference (triangles) for cooperative merging as a function of $p_2$. The parameters, other than $p_2$, are the same as in Fig. 3.

Fig. 6. (Color online) The difference between the total exiting and the number of merges as a function of $p_2$ for normal merging (diamonds) and cooperative merging (squares). The parameters, other than $p_2$, are the same as in Fig. 3.

Fig. 7. (a) Position and velocity of merges for cooperative (diamonds) and normal merging (squares). (b) Velocity *vs.* position of vehicles in lane 1 at $t = 500$ s for cooperative (diamonds) and normal merging (squares). The parameters are $p_1 = 1.0$, $p_2 =$

0.2, $n = 3$, $h_0 = 40$ m with a lane-2 offset of 1000 m, $d_{merge} = 100$ m, and $z_0 = -1000$ m. The limiting velocity is 30 m/s. The delay time is $t_d = 0.75$ s and $0.5$ s $< \tau_n < 1.0$ s.

Fig. 8. Velocity *vs.* position of vehicles in lane 1 at $t = 500$ s for cooperative merging for $d_{merge} = 100$ m (diamonds) and $d_{merge} = 300$ m (squares). The other parameters are the same as in Fig. 7.

Fig. 9. (a) Velocity *vs.* position at $t = 500$ s for both lanes. (b) Flow out of merge region measured at $x = 100$ m as a function of time. Cooperative merging is denoted by squares and normal merging by diamonds. The parameters are $p_1 = p_2 = 1$, $h_0 = 50$ m, $n = 3$, $d_{merge} = 300$ m, and $z_0 = -1000$ m.

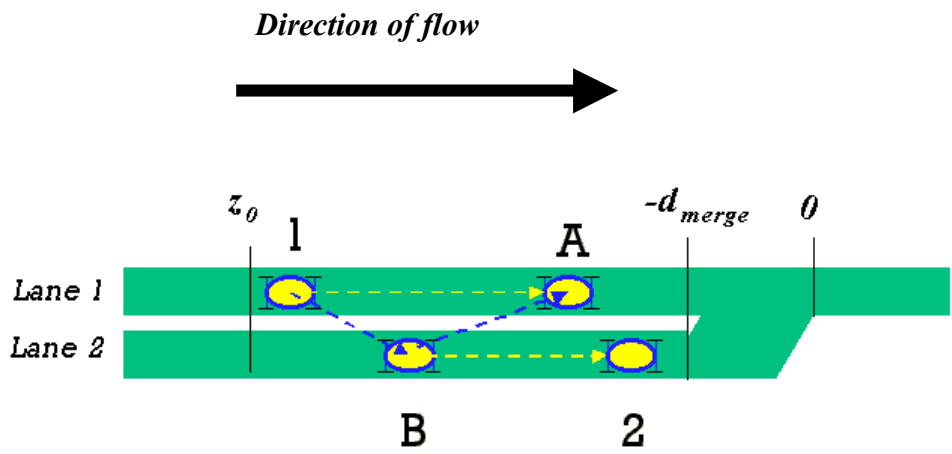

Fig. 1

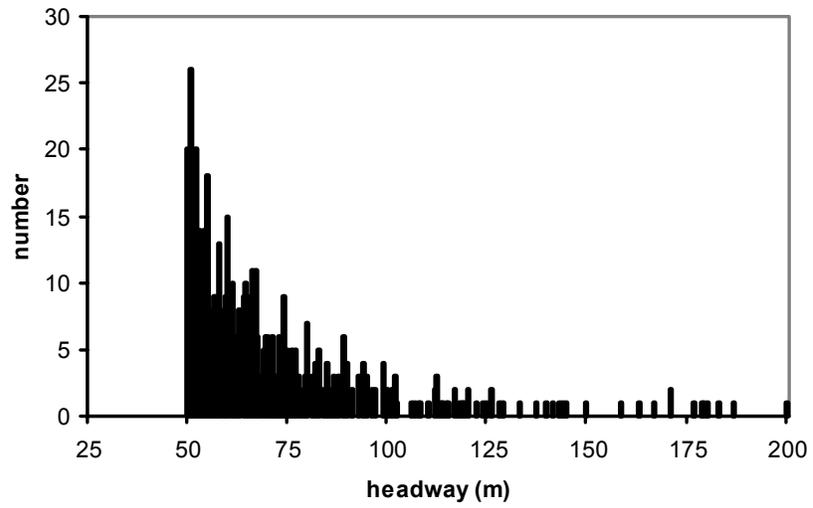

Fig. 2.

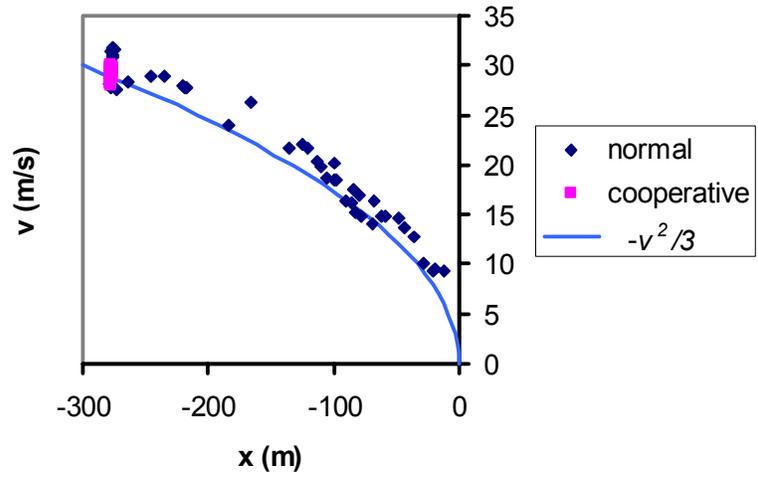

Fig. 3.

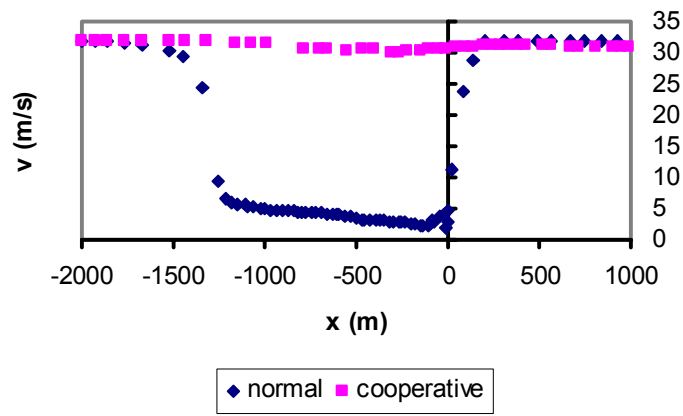

Fig. 4.

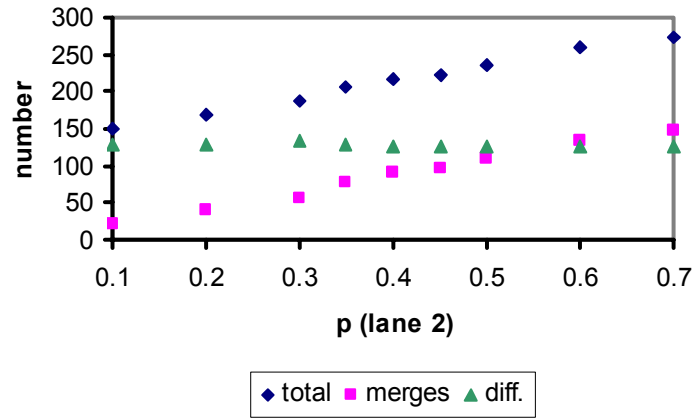

Fig. 5.

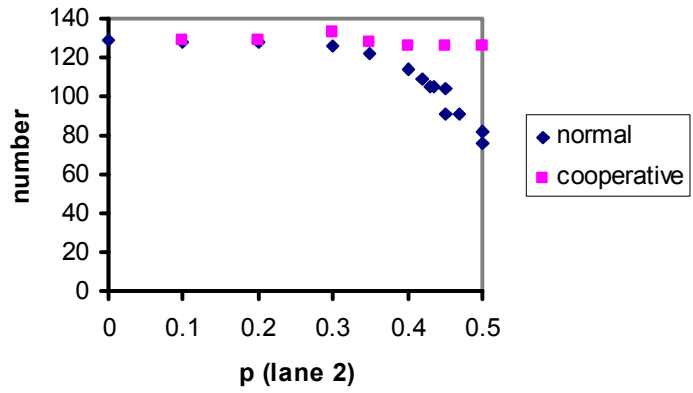

Fig. 6.

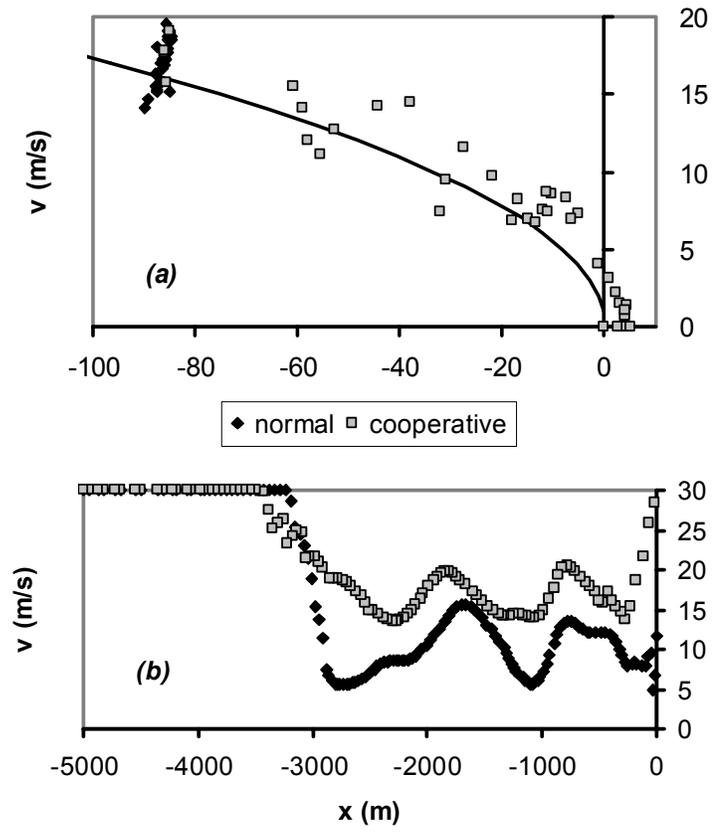

Fig. 7.

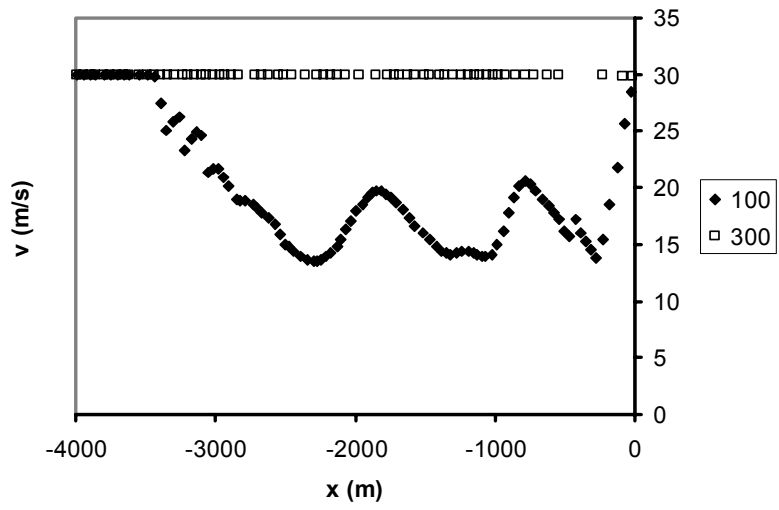

Fig. 8

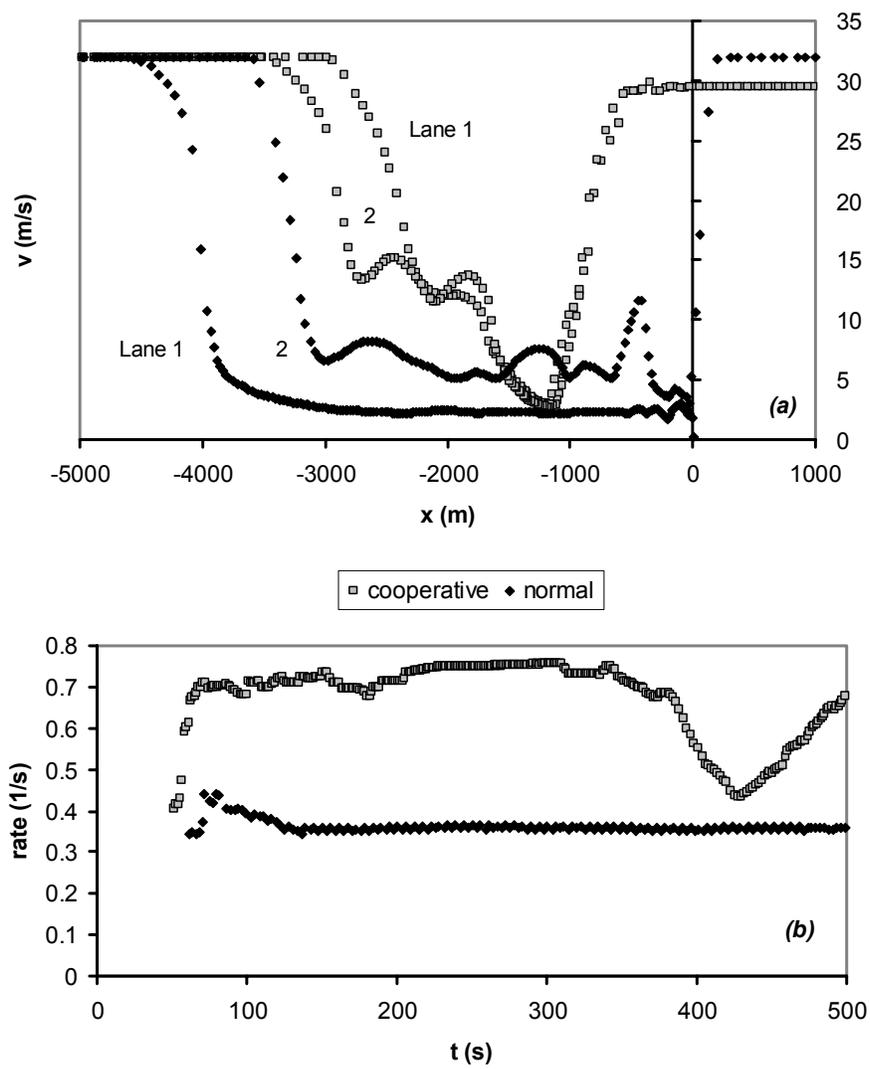

Fig. 9.